\documentclass[11pt,a4paper]{article}

\usepackage{amssymb}
\usepackage{amsmath}
\usepackage[pdftex]{graphicx}
\usepackage{graphics}
\usepackage{epsfig}
\usepackage{color}

\newcommand{\be}{\begin{equation}}
\newcommand{\ee}{\end{equation}}
\numberwithin{equation}{section}

\newtheorem{theor}{Theorem}[section]

\title{\bf The two lowest eigenvalues of the harmonic oscillator in the presence of a Gaussian perturbation}

\author{S. Fassari$^{1,2,3}$\footnote{silvestro.fassari@uva.es}, 
L.M. Nieto$^4$\footnote{luismiguel.nieto.calzada@uva.es},
F. Rinaldi$^{2,3}$\footnote{f.rinaldi@unimarconi.it}
\\ [2ex]
\footnotesize \sl $^1$Department of Higher Mathematics, ITMO University, S. Petersburg, Russian Federation\\
\footnotesize \sl $^2$CERFIM, PO Box 1132, CH-6601 Locarno, Switzerland\\
\footnotesize \sl $^3$Dipartimento di Fisica Nucleare, Subnucleare e delle Radiazioni, \\ 
\footnotesize \sl Univ. degli Studi Guglielmo Marconi,Via Plinio 44, I-00193 Rome, Italy\\
\footnotesize \sl $^4$Departamento de F\'{\i}sica Te\'{o}rica, At\'{o}mica y \'{O}ptica, and IMUVA,\\ 
\footnotesize \sl Universidad de Valladolid, 47011 Valladolid, Spain   
}

\begin{document}

\maketitle

\begin{abstract}
In this note we consider a one-dimensional quantum mechanical particle constrained by a parabolic well perturbed by a Gaussian potential. As the related Birman-Schwinger operator is trace class, the Fredholm determinant can be exploited in order to compute the modified eigenenergies which differ from those of the harmonic oscillator due to the presence of the Gaussian perturbation. By taking advantage of Wang's results on scalar products of four eigenfunctions of the harmonic oscillator, it is possible to evaluate quite accurately the two lowest lying eigenvalues as functions of the coupling constant $\lambda$.
\end{abstract}


\newpage

\section{Introduction}
As is well known, the harmonic oscillator is one of the very few solvable quantum models, that is to say its eigenfunctions and eigenvalues can be expressed analytically. As a consequence, any Quantum Mechanics textbook such as \cite{LA}, contains a chapter devoted to its detailed description.

This remarkable property has stimulated researchers to study various types of models involving perturbations of the harmonic oscillator over many decades. The interested reader can find a brief review of the literature on time independent perturbations of the harmonic oscillator in  \cite{Nano10} (see also \cite{RSII} and \cite{RSIV}).

Although the Birman-Schwinger principle was used in \cite{F 96} and \cite{FI97R} to investigate the spectral effects of a particular type of short range perturbation of the one-dimensional harmonic oscillator, namely a Lorentzian perturbation, the method is clearly applicable to any absolutely summable potential. Given our recent interest in various quantum models involving Gaussian perturbations (see \cite{Nano10}, \cite{MFRM}, \cite{FGNR}, \cite {AFGNR}), we have decided to make use of the above-mentioned principle in order to investigate the modifications of the discrete spectrum of the one-dimensional harmonic oscillator once a Gaussian perturbation is added to the Hamiltonian $H_0=\frac 1{2} \left(-\frac {d^2}{dx^2}+x^2\right)\geq \frac 1{2}$. 

\smallskip

Furthermore, we have been motivated to study such a model due to the lack of relevant contributions to the existing literature in theoretical/mathematical physics. As a matter of fact, to the best of our knowledge, the most relevant work on this model is to be found in the chemical literature, namely \cite{earl} (see also the previous contributions cited therein). As far as the physics literature is concerned, near the completion of the current work we happened to encounter a recent contribution \cite{2D} on the two-dimensional analogue of the model being investigated here. Since both works rely on different methods to compute the matrix elements related to the Gaussian perturbation, we thought it might be worth shedding further light on the model by taking advantage of some standard functional analytic tools. Although such tools might seem beyond the average mathematical knowledge of most physicists, they have been widely used by mathematical physicists in order to deal with various quantum mechanical perturbative problems.

\smallskip

Moving to the description of the contents of our paper and its findings, we start with some fundamental functional analytic preliminaries, the renowned Birman-Schwinger principle being the most important of them, that will be crucially exploited throughout our paper. 

In Section 3 we first consider the case of an attractive Gaussian perturbation. By realising that the divergence of the Birman-Schwinger kernel of our perturbed oscillator takes place only on a one-dimensional subspace, we find a relatively simple equation, the solution of which represents the energy of the perturbed ground state (its existence is guaranteed by the implicit function theorem). Finally, by means of an accuarte evaluation of the second order coefficient (the details of which are fully shown in the Appendix), we provide the approximation of the ground state energy up to the second order of the coupling constant.

In Section 4 the same technique will be adapted to achieve an equation leading to the calculation of the energy of the perturbed first excited state. Also in this case we explicitly write its approximation up to the second order of the coupling constant.

In Section 5 the findings obtained in the attractive case will be suitably modified to get also those for its repulsive counterpart.

Finally, Section 6 will include our final considerations while all the mathematical calculations required in the previous sections will be shown in full detail in the Appendix.

\section{Preliminaries} 
The  Schr\"odinger Hamiltonian for the one-dimensional harmonic oscillator perturbed by an attractive central Gaussian potential is given by:
\begin{equation}\label{1} 
H_\lambda= H_0-\lambda V(x)=H_0-\lambda e^{-x^2},\quad \lambda>0,
\end{equation} 
where 
\be\label{H_0}
H_0=\frac 1{2} \left(- \frac {d^2}{dx^2}+x^2\right)\geq \frac 1{2}
\ee
 is the Hamiltonian of the unperturbed harmonic oscillator. Hence, for any $ E<\frac 1{2}$, the equation determining the lowest eigenvalue (ground state energy) reads:
\begin{equation}\label{eigeq} 
\left[H_0-\lambda e^{-x^2}\right]\psi=E\psi \quad \Longleftrightarrow\quad 
\left[H_0- E  \right]\psi=\lambda e^{-x^2}\psi.
\end{equation}

\begin{figure}[h]
\centering
\includegraphics[width=9cm,keepaspectratio]{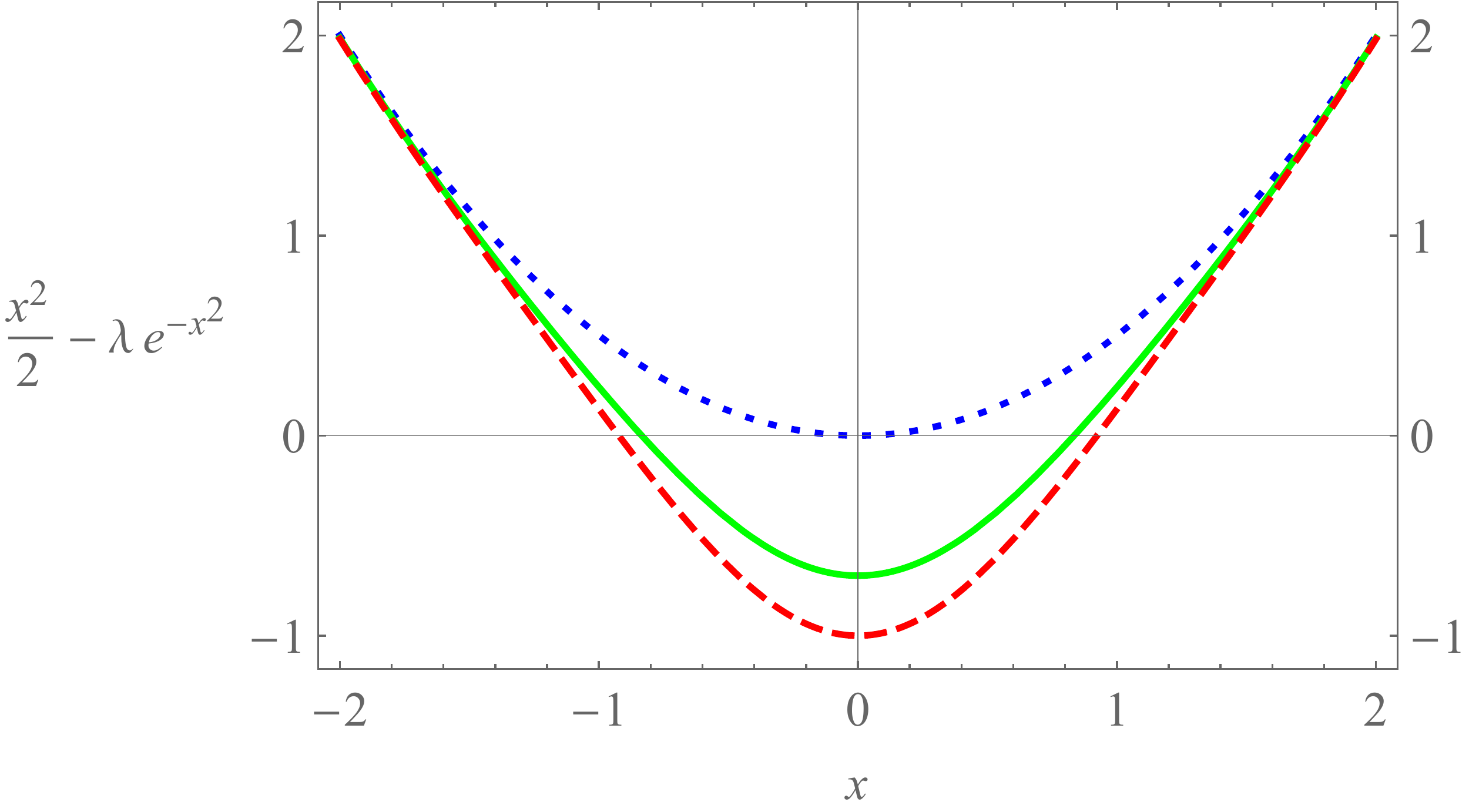}
\caption{The ``funnel" potential $\frac12 x^2-\lambda e^{-x^2}$: blue dotted line for $\lambda=0$ (harmonic oscillator), 
green solid line for $\lambda=0.7$, and red dashed line for $\lambda=1$.}
\label{funnel}
\end{figure}

The total potential has the shape of a curved ``funnel" potential (borrowing the terminology used in \cite{AnnPhys17}), as depicted in Figure~\ref{funnel}.

As shown in full detail in \cite{Nano10}, the above differential equation can be recast as the following integral equation
\begin{equation}\label{eigeq3} 
\chi=\lambda \left[H_0- E  \right]^{-\frac 1{2}}e^{-x^2}\left[H_0- E \right]^{-\frac 1{2}}\chi,
\end{equation}
with $\chi= \left[H_0- E   \right]^{\frac 1{2}}\psi$, taking account that the square root of the resolvent in $x$-space is a positive integral operator for any $E<\frac 1{2}$. The integral operator on the right hand side of the above equation reads:
\begin{equation}\label{eigeq3'} 
\lambda  \sum _{m=0}^{\infty } \sum _{n=0}^{\infty } \frac {| \psi_m \rangle (\psi_m , e^{-(\cdot)^2}  \psi_n)\langle \psi_n| }{(m+\frac 1{2}-E)^{1/2}(n+\frac 1{2}-E)^{1/2}},
\end{equation}
 where $\psi_n$ is the $n$-th normalised eigenfunction of the harmonic oscillator
 \be
 \label{harmonicoscillator}
\psi_n(x)=\frac1{\sqrt{2^n n! \sqrt{\pi}}}\ e^{-x^2/2}\, H_n(x), \quad \hbar=m=\omega=1,
\ee
$H_n(x)$ being the $n$-th Hermite polynomial.

Alternatively, by setting $\phi=e^{-x^2/2} \psi$, one gets instead a different integral equation, namely:
\begin{equation}\label{eigeq4} 
\phi=\lambda e^{-x^2/2}\left[H_0- E  \right]^{-1}e^{-x^2/2}\phi ,
\end{equation}
the integral operator on the right hand side being exèlicitly given by
\begin{equation}\label{eigeq4'} 
\lambda  \sum _{n=0}^{\infty } \frac {e^{-(\cdot)^2/2}|\psi_n \rangle\langle  \psi_n|e^{-(\cdot)^2/2}}{n+\frac 1{2}-E}.
\end{equation}

We remind the reader that the positive integral operator on the right hand side of  equation \eqref{eigeq4}  is the renowned  Birman-Schwinger operator, widely used in the literature on small perturbations of the Laplacian in the sense of quadratic forms, and that the two integral operators are isospectral (see \cite {KLA}, \cite {F 95}). The key result established in \cite{Nano10} is that, for any $E<\frac{1}{2}$, the positive isospectral integral operators $$\left(H_0-E \right)^{-1/2}e^{-x^2}\left (H_0- E  \right)^{-1/2}\quad \text{and}\quad  e^{-\frac {x^2}{2}} \left(H_0- E  \right)^{-1} e^{-\frac {x^2}{2}}$$ are trace class and their trace class norm is equal to
\begin{equation} \label{theor}
 \left|  \left| \lambda e^{-\frac {x^2}{2}} \left(H_0- E  \right)^{-1}e^{-\frac {x^2}{2}} \right| \right|_1=\lambda \left(\frac {\pi}{2} \right)^{\frac 1{2}} \frac{\Gamma \left( \frac 1{2}- E \right)}{\Gamma \left( 1- E \right)}.
\end{equation}

As an immediate consequence of the latter property, the Fredholm determinant (see \cite{RSIV}) can be exploited in order to determine the ground state energy of the perturbed Hamiltonian $H_\lambda$, so that the new ground state energy $E_0$ is the particular value of $E$  solving the equation
\begin{equation} \label{det}
\text{det} \left[1-\lambda e^{-\frac {x^2}{2}} \left(H_0- E  \right)^{-1}e^{-\frac {x^2}{2}}\right]=0.
\end{equation}

In the following sections we are going to use the above equation to determine the lowest lying eigenvalues of the Hamiltonian $H_\lambda$.

\section{The ground state energy of $H_\lambda$}

Given that the Gaussian potential is absolutely summable, the techniques used in \cite{F 96} and \cite{FI97R} can be exploited to get an accurate approximation of the ground state energy and those of lowest excited states.

By looking at \eqref{eigeq4'}, it is evident that, as $E\rightarrow \frac 1{2}$ from below, the Birman-Schwinger operator diverges positively only on the one-dimensional subspace spanned by the vector $ \psi_0(x)e^{-x^2/2}$. After setting
\begin{equation}\label{BSsub} 
M_E^{(0)}= e^{-(\cdot)^2/2}\left(\left[H_0- E  \right]^{-1}-\frac {|\psi_0 \rangle\langle  \psi_0|}{\frac 1{2}-E}\right)e^{-(\cdot)^2/2}= \sum _{n=1}^{\infty } \frac {e^{-(\cdot)^2/2}|\psi_n \rangle\langle  \psi_n|e^{-(\cdot)^2/2}}{n+\frac 1{2}-E}, 
\end{equation}
 equation \eqref{det} reduces to
\begin{equation} \label{tr}
\text{det} \left[1-\lambda \frac {e^{-(\cdot)^2/2}|\psi_0 \rangle\langle  \psi_0|e^{-(\cdot)^2/2}}{\frac 1{2}-E}\left[1-\lambda M_E^{(0)} \right]^{-1} \right]=0,
\end{equation}
 as follows from \cite{RSIV}, \cite{F 96} and \cite{FI97R}. Since the operator inside the above determinant has rank equal to one and $\text{det} \left[1+A\right]=1+\text{Tr}A$ for any rank one operator $A$, \eqref{tr} becomes
\begin{equation} \label{det''}
1=\lambda \text{Tr} \left(\frac {e^{-(\cdot)^2/2}|\psi_0 \rangle\langle  \psi_0|e^{-(\cdot)^2/2}}{\frac 1{2}-E}\left[1-\lambda M_E^{(0)}\right]^{-1} \right).
\end{equation}

The trace of the above rank one operator can be expressed explicitly so that the above equation becomes
\begin{equation} \label{tr'}
1=\frac {\lambda}{\frac 1{2}-E}  \left(e^{-(\cdot)^2/2}\psi_0, \left[1-\lambda M_E^{(0)} \right]^{-1} \psi_0 e^{-(\cdot)^2/2} \right),
\end{equation}
 which, in turn, after using \eqref{BSsub}, setting $E=\frac 1{2}-\epsilon_0,\epsilon_0>0$ and multiplying both sides of the above equation by the denominator of its rhs, we get:
\begin{equation} \label{tr'''}
\epsilon_0=\lambda \left(e^{-(\cdot)^2/2}\psi_0, \left[1-\lambda \sum _{n=1}^{\infty } \frac {e^{-(\cdot)^2/2}|\psi_n \rangle\langle  \psi_n|e^{-(\cdot)^2/2}}{n+\epsilon_0}\right ]^{-1} \psi_0 e^{-(\cdot)^2/2} \right).
\end{equation}

The positive trace class operator inside the square brackets is invertible on the subspace orthogonal to $\psi_0$ as long as 
\begin{equation} \label{norm}
\lambda  \left|  \left| \sum _{n=1}^{\infty } \frac {e^{-(\cdot)^2/2}|\psi_n \rangle\langle  \psi_n|e^{-(\cdot)^2/2}}{n+\epsilon_0}\right| \right|_{\infty} \leq \lambda  \left|  \left| \sum _{n=1}^{\infty } \frac {e^{-(\cdot)^2/2}|\psi_n \rangle\langle  \psi_n|e^{-(\cdot)^2/2}}{n}\right| \right|_{\infty} <1.
\end{equation}

As the operator is positive, its operator norm is equal to the supremum over all the unitary vectors $\psi$ of
\begin{equation} \label{norm'}
\lambda  \sum _{n=1}^{\infty }  \frac {\left(\psi_{n}, e^{-(\cdot)^2/2}  \psi \right)^2}{n}\leq \lambda  \sum _{n=1}^{\infty } \left(\psi_{n}, e^{-(\cdot)^2/2}  \psi \right)^2\leq \lambda  \sum _{n=0}^{\infty } \left(\psi_{n}, e^{-(\cdot)^2/2}  \psi \right)^2=\lambda \left|  \left|  e^{-(\cdot)^2/2}  \psi \right| \right|_2^2,
\end{equation}
the last equality following easily from the orthonormality of the eigenfunctions of the harmonic oscillator. Since the rhs of \eqref{norm'} is bounded by
\begin{equation} \label{norm''}
\lambda \left|  \left|  e^{-(\cdot)^2/2}  \psi \right|  \right|_2^2 \leq \lambda \left|  \left|  \psi \right|  \right|_2^2 =\lambda,
\end{equation}
which implies $\lambda \left|  \left| M_{1/2}^{(0)}\right|  \right|_{\infty}\leq \lambda$. Hence, for any  $\lambda  <1,$ the operator $1-\lambda M_{1/2}^{(0)}$ is invertible. A slightly more satisfactory condition is
\begin{equation} \label{lambda0}
\lambda  <\lambda_0 =\frac 1{\sqrt{2} \ln2} \approx 1.020,
\end{equation}
as follows from the exact evaluation of the trace class norm of the operator $\lambda  M_{1/2}^{(0)}$ shown in the Appendix~\ref{a1}.

The existence of a solution $\epsilon_0 (\lambda)>0$ of \eqref{tr'''} is guaranteed by the implicit function theorem, as can be seen by essentially mimicking the argument outlined in \cite{RSIV} (respectively \cite{F 96}) in the case of the corresponding equation for the operator $- \frac {d^2}{dx^2}+V(x), V\in C_0^{\infty}(-\infty,\infty), \int _{-\infty}^{\infty} V(x)dx \leq 0$ (resp. $- \frac {d^2}{dx^2}+x^2-\frac {\lambda}{g(1+gx^2)}$).
As was done in those references, a satisfactory approximation of the solution $\epsilon_0 (\lambda)>0$ (omitting the $o(\lambda^2)$-remainder) can be achieved by considering only the constant term and the linear one evaluated at $\epsilon=0$ in the Neumann expansion of $ \left[1-\lambda  M_E^{(0)} \right]^{-1}$, namely:
\begin{equation} \label{tr''''}
\epsilon_0(\lambda)=\lambda \sqrt{\pi} \left(\psi_0^2, \psi_0^2 \right) + \lambda^2 \pi  \sum _{n=1}^{\infty } \frac {\left(\psi_0, \psi_0^2 \psi_{2n} \right)^2}{2n},
\end{equation}
taking account of the fact that $\left(e^{-(\cdot)^2/2}\psi_0, e^{-(\cdot)^2/2}\psi_{2n+1} \right)=0, n \geq 0$ and that $e^{-x^2/2}= \pi^{1/4}\psi_0(x)$. Of course, as was to be expected, the rhs of \eqref{tr''''} is nothing else but the sum of the linear term and the quadratic one resulting from the perturbation series appearing in any Quantum Mechanics textbook (see, e.g., \cite{LA}). The scalar products on the rhs of \eqref{tr''''} can be computed explictly as a consequence of the results of \cite{Wa} (see also \cite{Nano10} and \cite{AFGNR}):
$$
\left(\psi_0^2, \psi_0^2 \right)=\frac {\psi_0^2(0)}{\sqrt{2}}=\frac 1{\sqrt{2\pi}}\quad\text{and} \quad\left(\psi_0, \psi_0^2 \psi_{2n} \right)^2=\frac {\psi_{2n}^2(0)}{2^{2n+1}\sqrt{\pi}} .
$$ 
Therefore, \eqref{tr''''} can be rewritten as follows
\begin{equation} \label{gseq}
\epsilon_0(\lambda)=\frac {\lambda}{\sqrt{2}} + \lambda^2 \frac {\sqrt{\pi}}{2} \sum _{n=1}^{\infty } \frac { \psi_{2n}^2(0)}{2^{2n+1}n}.
\end{equation}

As is well known, $ \psi_{2n}(x_0)$ behaves like $n^{-1/4}$ as $n \rightarrow +\infty$ for any fixed $x_0$ (see \cite{MS16,M14,FI94,FI96,FI97,FR12}), so that the strictly positive sequence inside the sum decays even more rapidly than exponentially as $n \rightarrow +\infty$, which guarantees the fast convergence of the positive series.  It is worth pointing out that the latter series contains $2^{2n}$ in the denominator of its sequence, differently from the series appearing in various models involving point perturbations of the harmonic oscillator (see \cite{MS16,M14,FI94,FI96,FI97,FR12,AFR,AFR2,AFRNano,AFRNano2,FGGN2,FGGNR}). As a result, we cannot expect to express the series in terms of a ratio of Gamma functions, as was done in the above-mentioned papers.

As will be shown in detail in the Appendix~\ref{a2}, 

\begin{equation} \label{gseq'}
\sum _{n=1}^{\infty } \frac { \psi_{2n}^2(0)}{2^{2n+1}n}=\frac {\ln (8-4\sqrt{3})}{\sqrt{\pi}}  .
\end{equation} 
Therefore, the approximation of the ground state energy up to the second order in the coupling constant is given by:
\begin{equation} \label{gseq''}
E_0(\lambda)=\frac 1{2}-\epsilon_0(\lambda)=\frac 1{2}-\frac {\lambda}{\sqrt{2}} - \lambda^2 \frac {\ln (8-4\sqrt{3})}{2}
\approx  0.5- 0.707\, \lambda - 0.035 \lambda^2 .
\end{equation}
A plot of this ground state energy as a function of $\lambda$ can be seen in Figure~\ref{0-1-states}.

\section{The energy of the first excited  state of $H_\lambda$}
By setting 
\begin{equation}\label{BSsub'} 
M_E^{(1)}= e^{-(\cdot)^2/2}\left(\left[H_0- E  \right]^{-1}-\frac {|\psi_1 \rangle\langle  \psi_1|}{\frac 1{2}-E}\right)e^{-(\cdot)^2/2}= \sum _{n\neq1}^{\infty } \frac {e^{-(\cdot)^2/2}|\psi_n \rangle\langle  \psi_n|e^{-(\cdot)^2/2}}{n+\frac 1{2}-E}, 
\end{equation}

\noindent  which, for any $E \in (\frac 1{2}, \frac 3{2})$, can be explicitly written as follows
\begin{equation}\label{BSsub''} 
M_E^{(1)}= \sum _{n=2}^{\infty } \frac {e^{-(\cdot)^2/2}|\psi_n \rangle\langle  \psi_n|e^{-(\cdot)^2/2}}{n+\frac 1{2}-E}-\frac {e^{-(\cdot)^2/2}|\psi_0 \rangle\langle  \psi_0|e^{-(\cdot)^2/2}}{E-\frac 1{2}}, 
\end{equation}

\noindent we can repeat all the steps of the previous section to get:
\begin{equation*} \label{tr1}
\epsilon_1=\lambda 
 \left(e^{-(\cdot)^2/2}\psi_1,
  \frac1{\displaystyle
  1+ \lambda\frac {e^{-(\cdot)^2/2}|\psi_0 \rangle\langle \psi_0|e^{-(\cdot)^2/2}}{1-\epsilon_1} 
  -\lambda  \sum _{n=2}^{\infty } \frac {e^{-(\cdot)^2/2}|\psi_n\rangle\langle \psi_n|e^{-(\cdot)^2/2}}{n-1+\epsilon_1}
} \
 \psi_1 e^{-(\cdot)^2/2} \right).
\end{equation*}

As is evident from the denominator, the operator fails to be invertible on the subspace orthogonal to that spanned by $\psi_0$ and $\psi_1$. By mimicking what was done in the previous section, we can get the following condition ensuring the existence of the inverse of the operator in the above denominator:
\begin{equation} \label{lambda1}
\lambda  <\lambda_1 =\frac 4{\sqrt{2}\left(1+2 \ln2 \right)} \approx 1.185.
\end{equation}

In perfect analogy with the result of the previous section, the solution $\epsilon_1(\lambda)$, whose existence is ensured by the implicit function theorem, can be satisfactorily approximated by
\begin{equation} \label{tr1'}
\epsilon_1(\lambda)=\lambda \sqrt{\pi} \left(\psi_1^2, \psi_1^2 \right) + \lambda^2 \pi  \left[\sum _{n=2}^{\infty } \frac {\left(\psi_1, \psi_0^2 \psi_{n} \right)^2}{n-1}-\left(\psi_1, \psi_0^2 \psi_0 \right)^2\right ],
\end{equation}
 which, taking account of the fact that $\left(\psi_1, \psi_0^2 \psi_{2n} \right)=0, n \geq 0$, reduces to 
\begin{equation} \label{tr1''}
\epsilon_1(\lambda)=\lambda \sqrt{\pi} \left(\psi_1^2, \psi_1^2 \right) + \lambda^2 \pi \sum _{n=1}^{\infty } \frac {\left(\psi_1, \psi_0^2 \psi_{2n+1} \right)^2}{2n}.
\end{equation}

Taking advantage of the results of \cite{Wa},
$$
\left(\psi_1^2, \psi_1^2 \right)=\frac {\Gamma (\frac 5{2})}{\sqrt{2} \pi}=\frac{3\sqrt{2}}{8\sqrt{\pi}}
\quad\text{and}
\quad\left(\psi_1, \psi_0^2 \psi_{2n+1} \right)=\frac {(n+1)\psi_{2(n+1)}^2(0)}{2^{2(n+1)}\sqrt{\pi}},
$$
then  \eqref{tr1''} reads
\begin{equation} \label{tr1'''}
\epsilon_1(\lambda)=\lambda \frac {3\sqrt{2}}{8} + \frac {\lambda^2 \sqrt{\pi}}{2} \sum _{n=1}^{\infty } \frac {(n+1)\psi_{2(n+1)}^2(0)}{2^{2(n+1)}n}=
 \frac {3\sqrt{2}}{8} \,  \lambda + \frac {2\sqrt{3}-3\left[1-\ln(8-4\sqrt{3})\right]}{24}\,  \lambda^2,
\end{equation} 
as is shown in the Appendix~\ref{a3}. 

\begin{figure}[htb]
\centering
\includegraphics[width=8cm,keepaspectratio]{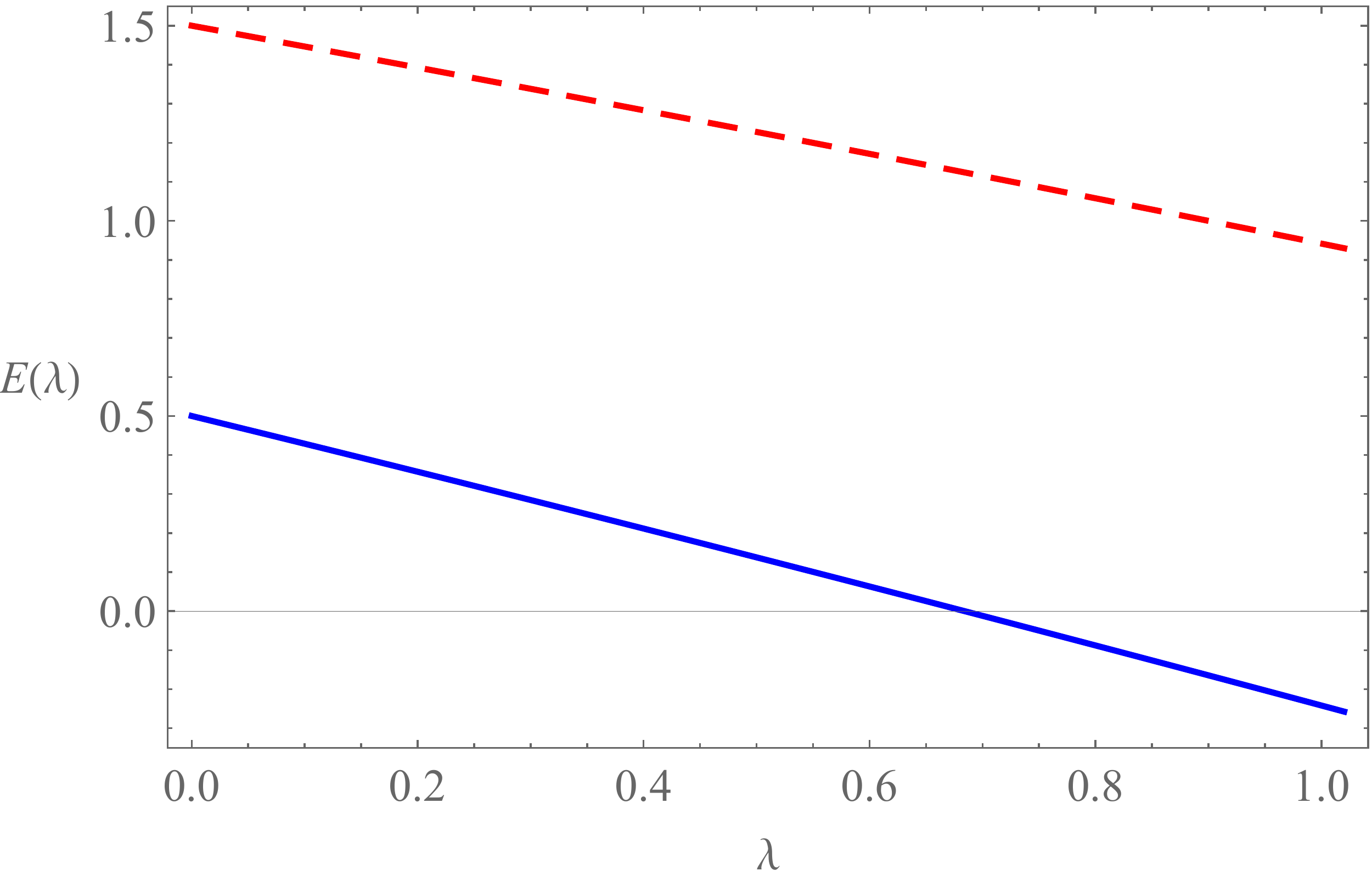}
\caption{The ground state energy $E_0(\lambda)$ (blue solid line) and the first excited state energy $E_1(\lambda)$ (red dashed line) of $H_\lambda$ as functions of the coupling constant $0\leq\lambda<\lambda_0 =\frac 1{\sqrt{2} \ln2} \approx 1.020$ from equations \eqref{gseq''} and \eqref{energyfirstexcited}.}
\label{0-1-states}
\end{figure}

Therefore, the approximation of the first excited state energy, up to the second order in the coupling constant, is given by:
\be\label{energyfirstexcited}
E_1(\lambda)=\frac32-\epsilon_1(\lambda) \approx  1.5-0.530\, \lambda -0.028\, \lambda^2 .
\ee
A plot of this first state energy as a function of $\lambda$ can be seen in Figure~\ref{0-1-states}. By comparing the two invertibility conditions \eqref{lambda0} and \eqref{lambda1}, it is clear that the joint plot of the two lowest eigenvalues of $H_\lambda$ may be drawn only up to $\lambda_0 =\frac 1{\sqrt{2} \ln2} \approx 1.020$.

\section{The two lowest energy levels in the repulsive case}
So far we have only considered the Hamiltonian $H_\lambda$, corresponding to an attractive Gaussian potential. However, taking advantage of what has been achieved in the attractive case, it is almost immediate to extend the results to the Hamiltonian  $H_{-\lambda}$ with a repulsive Gaussian potential. The total potential leads to a double well, as shown below in Figure~\ref{mexhat}, when $\lambda>\frac12$.
\begin{figure}[htb]
\centering
\includegraphics[width=9cm,keepaspectratio]{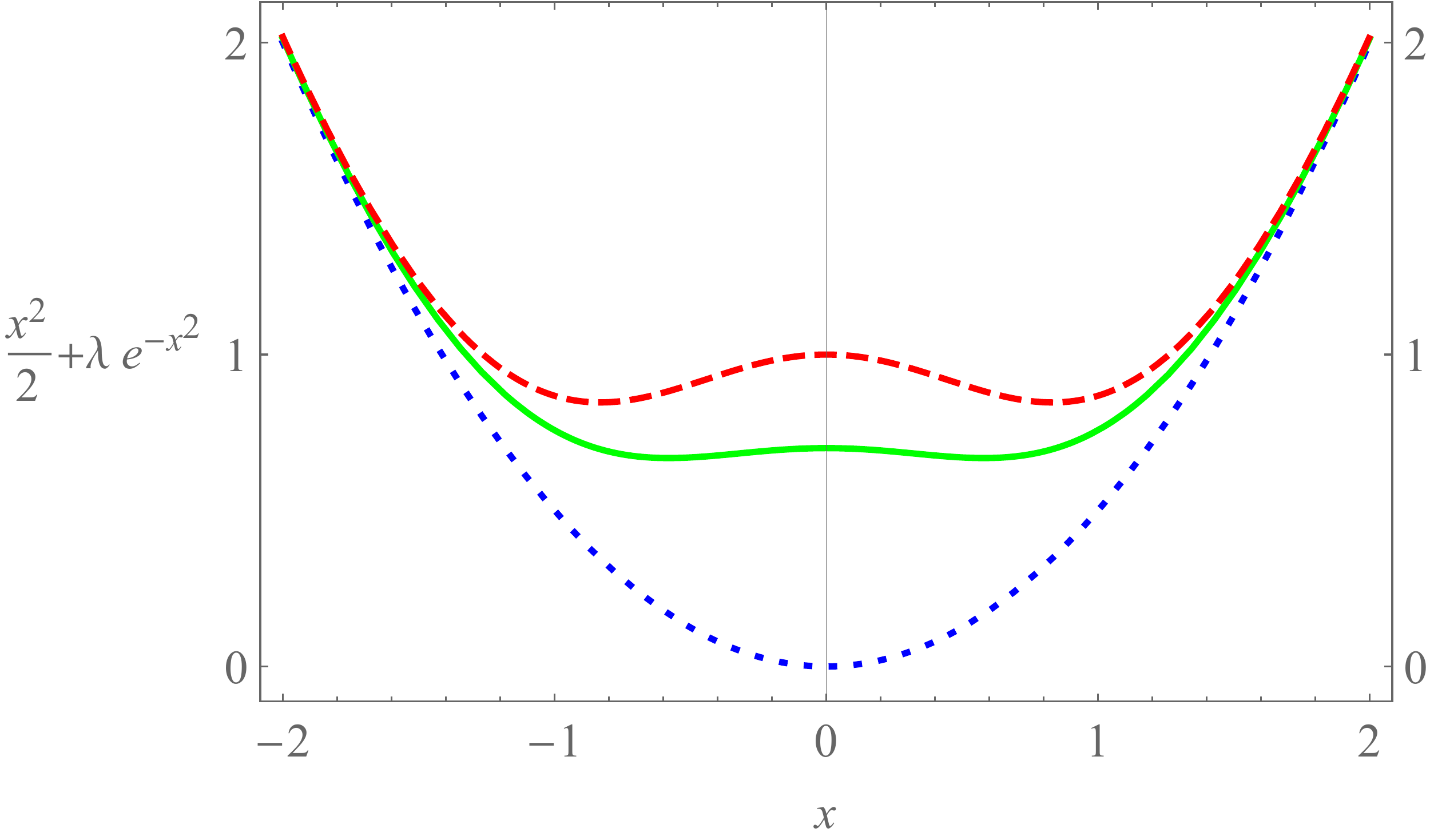}
\caption{The double well or ``mexican hat" potential $\frac12 x^2+\lambda e^{-x^2}$ when $\lambda>\frac12$: blue dotted curve for $\lambda=0$ (harmonic oscillator), green solid curve for $\lambda=0.7$, and red dasehd curve for $\lambda=1$.}
\label{mexhat}
\end{figure}

By replacing $\lambda$ with its opposite and $E_0=\frac 1{2}-\epsilon_0,\epsilon_0>0$ with $E_0=\frac 1{2}+\epsilon_0,\epsilon_0>0$, \eqref{tr'''} becomes:
\begin{equation} \label{rep1}
\epsilon_0=\lambda \left(e^{-(\cdot)^2/2}\psi_0, \left[1+\lambda \sum _{n=1}^{\infty } \frac {e^{-(\cdot)^2/2}|\psi_n \rangle\langle  \psi_n|e^{-(\cdot)^2/2}}{n-\epsilon_0}\right ]^{-1} \psi_0 e^{-(\cdot)^2/2} \right).
\end{equation}

It is worth stressing that, differently from the attractive case, the operator inside is fully invertible for any $\epsilon_0 \geq 0$ without any restriction on the coupling constant $\lambda>0$ since the lower bound of its spectrum is evidently equal to one. Therefore, in the repulsive case we get:
\begin{equation} \label{rep2}
\epsilon_0(\lambda)=\frac {\lambda}{\sqrt{2}} -  \lambda^2 \frac {\ln (8-4\sqrt{3})}{2},
\end{equation}
leading to the ground state energy
\begin{equation} \label{rep3}
E_0(\lambda)=\frac 1{2}+\frac {\lambda}{\sqrt{2}} -  \lambda^2 \frac {\ln (8-4\sqrt{3})}{2},
\end{equation}

As far as the first excited state is concerned, after replacing $\lambda$ with its opposite and $E_1=\frac 3{2}- \epsilon_1,\epsilon_1>0$ with $E_1=\frac 3{2}+ \epsilon_1,\epsilon_1>0$, the equation determining the first antisymmetric energy level is:
\begin{equation*} \label{rep4}
\epsilon_1=\lambda 
 \left(e^{-(\cdot)^2/2}\psi_1,
  \frac1{\displaystyle
  1- \lambda\frac {e^{-(\cdot)^2/2}|\psi_0 \rangle\langle \psi_0|e^{-(\cdot)^2/2}}{1+\epsilon_1} 
  +\lambda  \sum _{n=2}^{\infty } \frac {e^{-(\cdot)^2/2}|\psi_n\rangle\langle \psi_n|e^{-(\cdot)^2/2}}{n-1-\epsilon_1}
} \
 \psi_1 e^{-(\cdot)^2/2} \right) .
\end{equation*} 

\begin{figure}[h]
\centering
\includegraphics[width=8cm,keepaspectratio]{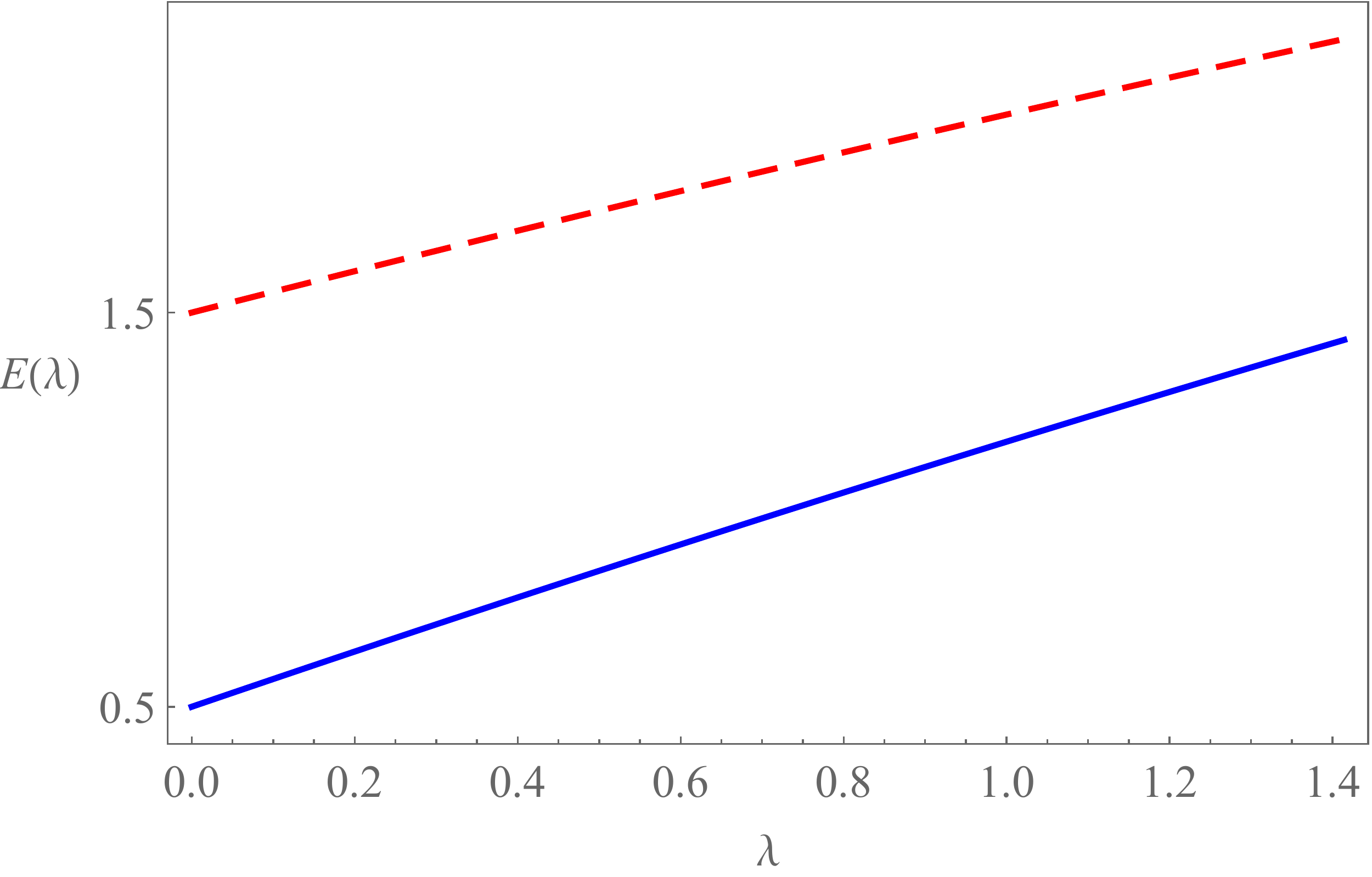}
\caption{The ground state energy $E_0(\lambda)$ (blue solid line) and the first excited state energy $E_1(\lambda)$ (red dashed line) of the double well or ``mexican hat" potential $\frac12 x^2+\lambda e^{-x^2}$ as functions of the coupling constant $0\leq\lambda< \sqrt{2}$, from equations \eqref{rep3} and \eqref{rep6}.}
\label{mexhat-states}
\end{figure}

It is crucial to notice that, differently from the attractive case, the operator in the denominator fails to be invertible only on the one-dimensional subspace spanned by $\psi_0$. The ensuing restriction on the coupling constant is clearly:
\begin{equation} \label{rep5}
\lambda <\frac 1{\sqrt{\pi} \left(\psi_0, \psi_0^2 \psi_0 \right)}=\sqrt{2}.
\end{equation}
Finally, the first excited energy level approximated to the second order in $\lambda$ is:
\begin{equation} \label{rep6} 
E_1(\lambda)=\frac 32+ \frac {3\sqrt{2}}{8} \,  \lambda - \frac {2\sqrt{3}-3\left[1-\ln(8-4\sqrt{3})\right]}{24}\,  \lambda^2
\end{equation}
the plot of which is provided in Figure~\ref{mexhat-states}.

\section{Final remarks}
 In this note we have investigated in detail how a Gaussian (repulsive or attractive) perturbation affects the two lowest energy levels of the harmonic oscillator, a perturbation previously investigated in the chemical literature. In particular, by exploiting two well-known functional analytic tools, precisely the Birman-Schwinger principle and the Fredholm determinant for trace class (also known as nuclear) operators, we have been able to provide explicitly their approximation up to the second order of the coupling constant. Although the method can be conceptually extended also to higher levels, we have restricted our analysis to the ground state energy and that of the first antisymmetric state due to the increasing computational complexity required.

By taking account of \eqref{gseq''} (respectively \eqref{rep3}) and \eqref{energyfirstexcited} (resp. \eqref{rep6}) and  looking at their visualisation in Fig.~\ref{0-1-states} (resp. Fig.~\ref{mexhat-states}), we can notice that in the attractive (resp. repulsive) case the first ionisation energy expands slightly more (shrinks slightly less) than linearly due to the smallness of the negative quadratic correction.

Although it is conceptually possible to get third order approximations in $\lambda$ for $E_0(\lambda)$ and $E_1(\lambda)$ by considering even the quadratic term in the Neumann expansion of $\left[1-\lambda M_ E^{(0)}\right]^{-1}$ and $\left[1-\lambda M_ E^{(1)}\right]^{-1}$, the increasing computational complexity of the task has made us decide to put off achieving that goal.

Motivated to a certain extent by \cite{2D}, it might be worth pointing out that the method can be extended to the 2D/3D analogues of our model, even though, given that the 2D/3D counterparts of our Birman-Schwinger operator are no longer nuclear but only Hilbert-Schmidt integral operators (see \cite{RSII,FI96,AFRNano,AFRNano2,FGGN2,FGGNR}),  the Fredholm determinant will have to be replaced by its modified version used for such operators (see \cite{GeKi}). Work in this direction is in progress.

\section*{Acknowledgements}
S. Fassari's contribution to this work has been made possible by the financial support granted by the Government of the Russian Federation through the ITMO University Fellowship and Professorship Programme. S. Fassari would like to thank Prof. Igor Yu. Popov and the entire staff at the Departament of Higher Mathematics, ITMO University, St. Petersburg for their warm hospitality throughout his stay. L.M. Nieto acknowledges partial financial support to 
Junta de Castilla y Le\'on and FEDER (Projects VA137G18 and BU229P18).

\appendix

\section{Appendix: proofs of some of the previous results}
In this appendix we wish to provide the reader with the mathematical details of some results used throughout the article.

\subsection{Calculation of the optimal range of admissible $\lambda$ ensuring the invertibility of $\lambda  M_{1/2}^{(0)}$}\label{a1}

\begin{theor} 
The operator $\left[1-\lambda M_{1/2}^{(0)}   \right]^{-1}$
$$
M_{1/2}^{(0)}=\sum _{n=1}^{\infty } \frac {e^{-(\cdot)^2/2}|\psi_n \rangle\langle  \psi_n|e^{-(\cdot)^2/2}}{n},
$$ 
exists for any positive $\lambda  <\frac 1{\sqrt{2} \ln2} \approx 1.020$.
\end{theor} 

\noindent
{\bf Proof.} As $M_{1/2}^{(0)}>0$, its trace class norm is exactly its trace, so that
\begin{equation} \label{M 1/2}
\text{tr} [M_{1/2}^{(0)}]=\sum _{m=0}^{\infty }\sum _{n=1}^{\infty } \frac {(\psi_m,e^{-(\cdot)^2/2}\psi_n)^2}{n}=\sum _{n=1}^{\infty } \frac { \left|  \left|  e^{-(\cdot)^2/2}\psi_n  \right|  \right|_2^2}{n},
\end{equation}

\noindent as follows easily from the orthonormality of the eigenfunctions of the harmonic oscillator. Since $$ \left(\psi_n,e^{-(\cdot)^2}\psi_n \right)=\sqrt {\pi}\left(\psi_n, \psi_0^2 \psi_n \right)=\sqrt {\pi} \frac {\psi_{2n}^2(0)}{\sqrt{2}},$$
the rhs of \eqref{M 1/2} is equal to
\begin{equation} \label{M 1/2'}
\sqrt {2 \pi} \sum _{n=1}^{\infty } \frac {\psi_{2n}^2(0)}{2n}=\sqrt {2 \pi} \lim_{\epsilon \rightarrow 0_{+}} \left[ \sum _{n=0}^{\infty } \frac {\psi_{2n}^2(0)}{2n+\epsilon}  -  \frac {\psi_0^2(0)}{\epsilon} \right].
\end{equation}

By mimicking what was done in \cite{FI94,FI96,FI97}, the rhs of \eqref{M 1/2'} can be written as
\begin{equation} \label{M 1/2''}
\sqrt {2 \pi} \lim_{\epsilon \rightarrow 0_{+}} \left[\left(H_0-\frac 1{2}+\epsilon \right)^{-1}(0,0) -  \frac 1{\sqrt{\pi}\epsilon} \right]=\sqrt {2 \pi} \lim_{\epsilon \rightarrow 0_{+}} \left[ \frac 1{\sqrt{\pi}}\int_0^1 \frac {s^{\epsilon-1}}{(1-s^2)^{1/2}}ds -  \frac 1{\sqrt{\pi}\epsilon} \right].
\end{equation}

After expressing the second term inside the square brackets as an integral and simplifying the integrand, the latter limit becomes
\begin{equation} \label{M 1/2'''}
\sqrt {2} \lim_{\epsilon \rightarrow 0_{+}} \int_0^1 \frac {s^{\epsilon-1}\left[1-(1-s^2)^{1/2}\right]}{(1-s^2)^{1/2}}ds=\sqrt {2} \lim_{\epsilon \rightarrow 0_{+}} \int_0^1 \frac {s^{\epsilon+1}}{(1-s^2)^{1/2}\left[1+(1-s^2)^{1/2}\right]}ds.
\end{equation}

As we may perform the limit inside the integral, the above rhs becomes
\begin{equation} \label{M 1/2''''}
\sqrt {2} \int_0^1 \frac {s}{(1-s^2)^{1/2}\left[1+(1-s^2)^{1/2}\right]}ds=-\sqrt {2} \int_0^1 \frac {d(1-s^2)^{1/2}}{ds}\frac 1{\left[1+(1-s^2)^{1/2}\right]}ds,
\end{equation}
which, after setting $y=(1-s^2)^{1/2}$, gets transformed into
\begin{equation} \label{M 1/2''''}
\sqrt {2} \int_0^1 \frac 1{1+y}dy=\sqrt {2} \left[ \ln (1+y) \right]_0^1=\sqrt {2} \ln2.
\end{equation}
Therefore, $\left|  \left| M_{1/2}^{(0)}  \right|  \right|_1=\sqrt {2} \ln2$. Then, for any $\lambda  <\frac 1{\sqrt{2} \ln2}$, we have  $\lambda \left|  \left| M_{1/2}^{(0)}  \right|  \right|_1 <1$, which ensures the existence of $\left[1-\lambda M_{1/2}^{(0)}   \right]^{-1}$.

Before closing this subsection, it might be worth noting that the same result could have been achieved by expressing the integral on the rhs of \eqref{M 1/2''} as a ratio of values of the Gamma function, as was done in \cite{AFR,AFRNano,AFRNano2,FGGN2}.

\subsection{Proof of the expression for $\epsilon_0(\lambda)$ given in equation \eqref{gseq'}}\label{a2}

\begin{theor} 
The following positive series converges and the sum is
\begin{equation} \label{theor}
\sum _{n=1}^{\infty } \frac { \psi_{2n}^2(0)}{2^{2n+1}n}=\frac {\ln (8-4\sqrt{3})}{\sqrt{\pi}}.
\end{equation} 

\end{theor}

\noindent
{\bf Proof.}
First of all, we notice that:
\begin{equation} \label{series2}
\sum _{n=1}^{\infty } \frac { \psi_{2n}^2(0)}{2^{2n+1}n}
=\sum _{n=1}^{\infty } \frac { \psi_{2n}^2(0)}{2^{2n+1}} \int_0^{1} x^{n-1}dx
=\int_0^{1} \left[ \sum _{n=1}^{\infty } \frac { \psi_{2n}^2(0)}{2^{2n+1}}\,  x^{n-1} \right] dx
=\int_0^{1} \left[ \sum _{n=1}^{\infty } \psi_{2n}^2(0) \left(\frac{x}4\right)^n \right] \frac{dx}{2x}.
\end{equation}
\smallskip

As follows in a rather straightforward manner from the definition of the normalised eigenfunctions of the harmonic oscillator \eqref{harmonicoscillator} (see \cite {MS16,FI94,FGGNR}), 
$$
\psi_{2n}(0)= \frac{H_{2n}(0)}{\sqrt{2^{2n} (2n)! \sqrt{\pi}}},
$$
and as is well known (see \cite{RSII,Arf})
$$
H_{2n}(0)= (-1)^n \frac{(2n)!}{n!},
$$

so that the series inside the square brackets on the rhs of \eqref{series2} becomes
\begin{equation} \label{series3}
\sum _{n=1}^{\infty } \psi_{2n}^2(0) \left(\frac{x}4\right)^n  =
\sum _{n=1}^{\infty }    \frac { (2n)!}{\sqrt{\pi}\,  2^{2n}\, (n!)^2}      \left(\frac{x}4\right)^n  =
\frac1{\sqrt{\pi}} \sum _{n=1}^{\infty }    \frac { (2n)!}{(n!)^2}      \left(\frac{x}{2^4} \right)^n  
= \frac1{\sqrt{\pi}} \left(  \frac{2}{\sqrt{4-x}}-1\right) .
\end{equation}
Going with this result to the rhs of \eqref{series2} and integrating from 0 to 1, we get:
\begin{equation} \label{series6}
\sum _{n=1}^{\infty } \frac { \psi_{2n}^2(0)}{2^{2n+1}n}
= \frac1{2\sqrt{\pi}}  \int_0^{1} \left(  \frac{2}{\sqrt{4-x}}-1\right) \frac{dx}{x}= \frac{\ln (8-4\sqrt{3})}{\sqrt{\pi}},
\end{equation}
 which completes the proof of our claim.

\subsection{Sum of the series  in equation  \eqref{tr1'''}}\label{a3}

\begin{theor} 
The following positive series converges and the sum is
\begin{equation} \label{Theor}
\sum _{n=1}^{\infty } \frac {(n+1)\psi_{2(n+1)}^2(0)}{2^{2(n+1)}n}=\frac {2\sqrt{3}-3\left(1-\ln [8-4\sqrt{3}] \right)}{12\sqrt{\pi}}.
\end{equation}
\end{theor} 

\noindent
{\bf Proof.}
The proof is similar to the previous one: using the eigenfunctions  \eqref{harmonicoscillator}
\be
\psi_{2(n+1)}(0)= \frac{H_{2(n+1)}(0)}{\sqrt{2^{2(n+1)} (2n+2)! \sqrt{\pi}}},
\ee
and the fact that
$$
H_{2(n+1)}(0)= (-1)^{n+1} \frac{(2n+2)!}{(n+1)!},
$$

it is straightforward to deduce the following
\begin{eqnarray*} 
&& \sum _{n=1}^{\infty } \frac {(n+1)\psi_{2(n+1)}^2(0)}{2^{2(n+1)}n}=
\sum _{n=1}^{\infty } \frac {\psi_{2(n+1)}^2(0)}{2^{2(n+1)}}\, (n+1)
 \int_0^{1} x^{n-1}dx\\
 && \quad =\int_0^{1} \left[ 
\sum _{n=1}^{\infty } \frac {\psi_{2(n+1)}^2(0)}{2^{2(n+1)}}\, (n+1)\,  x^{n} 
\right] \frac{dx}x
=\int_0^{1} \frac{d}{dx }\left[ 
\sum _{n=1}^{\infty } \frac {\psi_{2(n+1)}^2(0)}{2^{2(n+1)}}\,  x^{n+1} 
\right] 
\frac{dx}{x}\\
&&\quad 
=\int_0^{1} \frac{d}{dx }\left[ 
\sum _{n=1}^{\infty }  \psi_{2(n+1)}^2(0)\, \left(\frac{x}{2^2}\right)^{n+1}
\right] 
\frac{dx}{x}
 =\int_0^{1} \frac{d}{dx }\left[  
\sum _{n=1}^{\infty }  
\frac{(2n+2)!}{2^{2(n+1)} (n+1)!^2 \sqrt{\pi}}
\left(\frac{x}{2^2}\right)^{n+1} \right] 
\frac{dx}{x}\\
&&\quad 
=\frac1{\sqrt{\pi}} \int_0^{1} \frac{d}{dx }\left[  
\sum _{n=1}^{\infty }  
\frac{(2n+2)!}{ (n+1)!^2 }
\left(\frac{x}{2^4}\right)^{n+1} \right] 
\frac{dx}{x}
=\frac1{\sqrt{\pi}} \int_0^{1} \frac{d}{dx }\left[  
\frac2{\sqrt{4-x}}-1-\frac{x}8
 \right] 
\frac{dx}{x}\\
&&\quad = \frac1{12\sqrt{\pi}}
\left(
2\sqrt{3}-3\left[1-\ln (8-4\sqrt{3}) \right]
\right),
\end{eqnarray*}
which completes our proof.


\begin{thebibliography}{99}
\bibitem{F 96} Fassari S 1996, A note on the eigenvalues of the Hamiltonian of the harmonic oscillator perturbed by the potential $\frac {\lambda x^2}{1+gx^2}$, Rep. Math. Phys. {\bf 37} (2) 283-293
\end{thebibliography}
\end{document}